\documentclass[12pt]{iopart}
\usepackage{graphicx,iopams,setstack}
\begin{document}
\title{Inhomogeneous Anisotropic Passive Scalars}

\author{M Martins Afonso\dag\ddag\ and M Sbragaglia\S}

\address{\dag\ INFM-Department of Physics, University of Genova, via Dodecaneso 33, I-16146 Genova, Italy}
\address{\ddag\ INFN, Genova Section, via Dodecaneso 33, I-16146 Genova, Italy}
\address{\S\ Department of Physics and INFN, University of Roma "Tor Vergata", via della Ricerca Scientifica 1, I-00133 Roma, Italy}

\ead{marcomar@fisica.unige.it}
\begin{abstract}
We investigate the behaviour of the two-point correlation function in the context of passive scalars for non homogeneous, non isotropic forcing ensembles. Exact analytical computations can be carried out in the framework of the Kraichnan model for each anisotropic sector. We will focus our attention on the isotropic sector with isotropic forcing in order to obtain a description of the influence of purely inhomogeneous contributions. It is shown how the homogeneous solution is recovered at separations smaller than an intrinsic typical lengthscale induced by inhomogeneities, and how the different Fourier modes in the centre-of-mass variable recombine themselves to give a ``beating'' (superposition of power laws) described by Bessel functions. The pure power-law behaviour is restored even if the inhomogeneous excitation takes place at very small scales.
\end{abstract}

\submitto{JOT}

\maketitle
\section{Introduction}
\label{sec:1}
Since the fundamental work by Kolmogorov in $1941$ \cite{frish},\cite{K41} one of the key points for many theoretical achievements in turbulence research is the (statistical) restoration of homogeneity and isotropy of fluctuations at small scales \cite{sree},\cite{arneo}.
It is however impossible to provide a proper and consistent description of a great variety of systems where ``non idealized'' fluctuations are still alive. The characterization of the emission of a tracer from point-like sources, the study of scalar concentrations along channels with non-homogeneous boundaries, or in those systems whose large scales are driven by strong shears, are remarkable examples in which strong anisotropies \cite{Kur},\cite{Shen} and inhomogeneities \cite{Dhr} must be taken into account in order to obtain a correct picture of the statistical properties of those systems.\\
In the last decade a consistent progress in the development of a systematic analysis to separate isotropic fluctuations from the anisotropic ones in real turbulent flows and turbulent transport \cite{procaccia},\cite{bif1},\cite{bif2} has been carried out. It has been understood how to treat and face systems whose rotational symmetries are broken by the presence of an external forcing inducing anisotropic contributions. In particular, the study of a simplified model for passive scalar advection, known as Kraichnan model \cite{kraic1},\cite{kraic2}, has provided a clear understanding of the statistical properties in all anisotropic sectors of the scalar fluctuations. Indeed, closed equations for the equal-time correlation functions can be obtained: these are linear partial differential equations whose unforced solutions (also called zero modes \cite{gaw}) generally exhibit anomalous scaling. This is in contrast with the forced solutions that possess (non-anomalous) dimensional scaling. This has given the insight to explain the universality in the statistical framework. Indeed, the anomalous properties of small scale statistics result from a decoupling between the zero mode scaling and dimensional scaling, and the universality of these properties naturally emerges because the zero-mode scaling properties are independent on the forcing mechanism (see \cite{falk} for an exhaustive review).\\
In the present paper we formulate the concept of the possible small-scale homogeneity restoration by focusing on the two-point equal-time scalar correlation function for the Kraichnan advection model. The advecting velocity is still homogeneous and isotropic but this is not for the scalar injection mechanism which is supposed here to be neither isotropic nor homogeneous. As we will show, the inhomogeneous forcing induces a new lengthscale $\ell_q$ into the scalar dynamics, in terms of which quantitative conclusions on the persistence of small-scale inhomogeneity will be given. The aim of this investigation is twofold. Firstly, we want to show how the correct homogeneous limit can be restored going at separations (in the two-point scalar correlation function) smaller than $\ell_q$. Secondly, we want to give some analytical insights about the opposite physical situation represented by the presence of inhomogeneous fluctuations on scales of the same order of our separation. In the latter regime the pure power-law behaviour (homogeneous limit) is replaced by the ``beating'' in superposition of different power laws originating from the scalar inhomogeneities.\\
The paper is organized as follows: in section~\ref{sec:2} we formulate the general problem of anisotropic and inhomogeneous correlations. This problem will be studied analytically for the Kraichnan ensembles. In section~\ref{sec:3} the statistical description of the isotropic sector is provided with particular emphasis on some kinds of physically meaningful forcings. Conclusions follow in section~\ref{sec:4}.
\section{The two-point non homogeneous, non isotropic scalar correlation functions}
\label{sec:2}
\subsection{Basic equations}
Let us start with the equation for a passive scalar field transported by the velocity field ${\bi v}$:
\begin{equation} \label{partenza}
\partial_t\theta({\bi x}_1,t)+{\bi v}({\bi x}_1,t)\cdot\bnabla\theta({\bi x}_1,t)=\kappa\Delta\theta({\bi x}_1,t)+f({\bi x}_1,t)\;.
\end{equation}
The velocity field ${\bi v}$ is assumed incompressible, statistically homogeneous and 
isotropic, whereas the source term $f$ is a large-scale random forcing
which is not invariant under translations. Let us now specialize to the case of the Kraichnan ensemble \cite{kraic1},\cite{kraic2} where the velocity field is Gaussian, white in time, zero-mean and with two-point correlation function
\[\langle v_{\alpha}({\bi x}_1,t_1)v_{\beta}({\bi x}_2,t_2)\rangle=D_{\alpha\beta}({\bi x}_1-{\bi x}_2)\delta(t_1-t_2)\;.\]
The spatial behaviour is described by
\[D_{\alpha\beta}({\bi r})=D_0\delta_{\alpha\beta}-d_{\alpha\beta}({\bi r})\;,\]
with
\[d_{\alpha\beta}({\bi r})=D_1r^{\xi}\left[(d+\xi-1)\delta_{\alpha\beta}-\xi\frac{r_{\alpha}r_{\beta}}{r^2}\right]\]
for $r=|{\bi r}|$ smaller than the integral scale of the velocity field ($L_v$), above which $d_{\alpha\beta}({\bi r})$ saturates to an almost constant value whose order of magnitude is $D_1L_v^{\xi}$; consequently, since the correlation $D_{\alpha\beta}({\bi r})$ has to vanish for $r \to \infty$, the relation $D_0 \sim D_1L_v^{\xi}$ holds. $d$ is the space dimension ($\ge 2$) and $\xi$ is the scaling exponent, describing the degree of roughness present in the velocity field, lying in the open interval $(0,2)$. The forcing is assumed to be Gaussian, white in time, zero-mean and with two-point correlation function $\langle f({\bi x}_1,t_1)f({\bi x}_2,t_2)\rangle=F({\bi x}_1,{\bi x}_2)\delta(t_1-t_2)$. The steady-state equation for the two-point equal-time scalar correlation function $C({\bi x}_1,{\bi x}_2) \equiv \langle\theta({\bi x}_1,t)\theta({\bi x}_2,t)\rangle$ reads
\begin{equation} \label{equazione}
\fl d_{\alpha\beta}({\bi r})\frac{\partial^2 C}{\partial r_\alpha\partial r_\beta}\!+\!\frac{1}{4}[D_{\alpha\beta}({\bi r})\!+\!D_{\alpha\beta}({\bf 0})]\frac{\partial^2 C}{\partial z_{\alpha}\partial z_{\beta}}\!+\!2\kappa\frac{\partial^2 C}{\partial r_{\alpha}\partial r_{\alpha}}\!+\!\frac{\kappa}{2}\frac{\partial^2 C}{\partial z_{\alpha}\partial z_{\alpha}}\!+\!F({\bi r},{\bi z})=0\;,
\end{equation}
where ${\bi z}=({\bi x}_1+{\bi x}_2)/2$ is the centre of mass and ${\bi r}={\bi x}_1-{\bi x}_2$ is the relative separation. To obtain (\ref{equazione}) we have multiplied (\ref{partenza}) by $\theta({\bi x}_2,t)$, averaged over the stochastic fields ${\bi v}$ and $f$ (exploiting Furutsu-Novikov's rule on integration by parts \cite{Novikov}), symmetrized the resulting expression and eventually dropped the temporal derivative term. Fourier transforming in ${\bi z}$ and defining
\[\hat{C} \equiv \hat{C}({\bi r},{\bi q})=\int\rme^{\rmi{\bi q}\cdot{\bi z}}C({\bi r},{\bi z})\rmd{\bi z}\;,\]
\[\hat{F} \equiv \hat{F}({\bi r},{\bi q})=\int\rme^{\rmi{\bi q}\cdot{\bi z}}F({\bi r},{\bi z})\rmd{\bi z}\;,\]
we obtain, in the limit $\kappa \to 0$, the corresponding equation for $\hat{C}$:
\[d_{\alpha\beta}({\bi r})\frac{\partial^2\hat{C}}{\partial r_{\alpha}\partial r_{\beta}}-
\frac{1}{4}[D_{\alpha\beta}({\bi r})+D_{\alpha\beta}({\bf 0})]\,q_{\alpha}q_{\beta}\hat{C}
+\hat{F}({\bi r},{\bi q})=0\;.\]
At small separations $r \ll L_v$, since $D_1 \sim D_0L_v^{-\xi}$, $d_{\alpha\beta}({\bi r}) \sim D_0(r/L_v)^{\xi}$ is negligible with respect to $D_0$ and $D_{\alpha\beta} \simeq D_0\delta_{\alpha\beta}$. In the limit $r \ll L_v$ we can thus consider the simpler equation 
\[d_{\alpha\beta}({\bi r})\frac{\partial^2\hat{C}}{\partial r_{\alpha}\partial r_{\beta}}-
\frac{1}{2}D_0q^2\hat{C}+\hat{F}({\bi r},{\bi q})=0\;.\]
Generally, all the anisotropic (in the separation $r$) components of $\hat{F}({\bi r},{\bi q})$ may induce effects that are not globally invariant under rotations. It is thus better to consider the following decomposition \cite{ARAD}: 
\[\hat{C}({\bi r},{\bi q})=\sum_{l,m}\hat{C}_{l,m}(r,{\bi q})Y_{l,m}(\Omega)\;,\]
\[\hat{F}({\bi r},{\bi q})=\sum_{l,m}\hat{F}_{l,m}(r,{\bi q})Y_{l,m}(\Omega)\;,\]
with $\Omega$ denoting the solid angle associated with ${\bi r}$, and study the behaviour of the correlation function at separations $r$ smaller than the forcing correlation length $L$, where $\hat{F}_{l,m}(r,{\bi q}) \simeq \hat{F}_{l,m}(0,{\bi q})$. The equation in each anisotropic sector $(l,m)$ for the correlation function $\hat{C}_{l,m}=\hat{C}_{l,m}(r,{\bi q})$ reads
\begin{equation} \label{eq:1}
\fl r^{-(d-1)}\partial_rr^{d+\xi-1}\partial_r\hat{C}_l-\frac{(d+\xi-1)}{d-1}l(d-2+l)r^{-2}\hat{C}_l-\ell_q^{-(2-\xi)}\hat{C}_l+\phi_{{\bi q},l}=0
\end{equation}
where, because of the degeneration, we have dropped the dependence on the subscript $m$ and where we have introduced a rescaled forcing in the form $\phi_{{\bi q},l}=\hat{F}_l(0,{\bi q})/(d-1)D_1$. A new scale $\ell_q$, associated to the strength of the scalar inhomogeneities, has also been introduced. It measures the separation above which inhomogeneities become important (e.g. it can be derived through a dimensional-analysis balance between the first and the third term in (\ref{eq:1})) and it is defined as
\[\ell_q=\left[\frac{q^2D_0}{2(d-1)D_1}\right]^{-1/(2-\xi)}\;.\]
The Fourier scale $q^{-1}$ of the forcing inhomogeneities can be larger or of the same order of $L$. In the latter case we have $\ell_q/L \sim (L/L_v)^{\xi/(2-\xi)} \ll 1$ and the lengthscale $\ell_q$ lies inside the inertial range. The homogeneous case is recovered when $q^{-1} \gg L$ so as to insure $\ell_q/L \gg 1$. An important exception to the previous limitation on $q^{-1}$ is provided, in way of example, by the emission of a tracer from a point source, in which all $q$'s are excited with the same strength. This case is however quite peculiar because $L$ would be infinitesimal and $\ell_q$ would fall in the diffusive range, thus preventing us from taking the limit $\kappa \to 0$. In what follows we will always consider $\kappa \to 0$ and $L_v \to \infty$.
\subsection{General solution}
It is easy to verify that the general solution of (\ref{eq:1}), as a function of $r$ and $\ell_q$, is
\begin{equation} \label{eq:2}
\hat{C}_{l}(r;\ell_q)=\hat{C}_{l;{\rm part}}(r;\ell_q) +r^{-(d+\xi-2)/2}\left[B_1K_{\nu_l}(w)+B_2I_{\nu_l}(w)\right]\;,
\end{equation}
where $w=2(2-\xi)^{-1}(r/\ell_q)^{(2-\xi)/2}$ and $\nu_l=[(d+\xi-2)^2+4(d+\xi-1)l(d-2+l)/(d-1)]^{1/2}/(2-\xi)$. The constants $B_1$,$B_2$ are fixed by the boundary conditions and $K$,$I$ represent the Bessel functions \cite{GRAD} of complex argument whose behaviours for small arguments are $K_{\nu_l}(w) \sim w^{-\nu_l}$ and $I_{\nu_l}(w) \sim w^{\nu_l}$. The particular solution $\hat{C}_{l;{\rm part}}(r;\ell_q)$ of (\ref{eq:1}) can be found, for instance, exploiting the method of variation of constants \cite{couhil} which leads to 
\[\fl\hat{C}_{l;{\rm part}}(r;\ell_q)=\phi_{{\bi q},l}Ar^{(2-d-\xi)/2}\left[K_{\nu_l}(w)\int_0^w\rho^{\nu_0+1}I_{\nu_l}(\rho)\rmd\rho+I_{\nu_l}(w)\int_w^{\infty}\rho^{\nu_0+1}K_{\nu_l}(\rho)\rmd\rho\right]\;,\]
where $A=\ell_q^{(2+d-\xi)/2}((2-\xi)/2)^{\nu_0}$ and $\nu_0=(d+\xi-2)/(2-\xi)$. Regularity at $r=0$ imposes $B_1=0$ since $K_{\nu_l}(w) \sim w^{-\nu_l}$, while the term with $B_2$ is regular as $r \to 0$ since $I_{\nu_l}(w) \sim w^{\nu_l}$.\\ 
An exact solution can also be found for the values $r \gg L$. Indeed, in this case, we have $\hat{F}_l(r,{\bi q}) \simeq 0$, as the forcing correlation rapidly decreases for separations much greater than $L$, and an unforced equation arises:
\begin{equation} \label{eq:5}
r^{-(d-1)}\partial_rr^{d+\xi-1}\partial_r\hat{C}_l-\frac{(d+\xi-1)}{d-1}l(d-2+l)r^{-2}\hat{C}_l-\ell_q^{-(2-\xi)}\hat{C}_l=0\;.
\end{equation}
The solution of (\ref{eq:5}) reduces to the zero mode,
\begin{equation} \label{eq:6}
\hat{C}_l(r;\ell_q)=r^{-(d+\xi-2)/2}\left[B_3K_{\nu_l}(w)+B_4I_{\nu_l}(w)\right]\;,
\end{equation}
where regularity at $\infty$ imposes $B_4=0$ as $I_{\nu_{l}}(w) \sim w^{-1/2}\rme^w$ for $r \to \infty$, while the term with $B_3$ is regular because $K_{\nu_{l}}(w) \sim w^{-1/2}\rme^{-w}$ for $r \to \infty$.\\
The correlation of a passive scalar field in the presence of inhomogeneous fluctuations, whose characteristic length is $\ell_q$, can be thus computed in the limits of small ($r \ll L$) and large ($r \gg L$) separations with respect to the integral scalar scale $L$. We find a dependence on some unknown constants ($B_2$,$B_3$) that can be fixed upon boundary conditions. To provide a clear and instructive example we can assume a forcing whose correlation function is a step function in $r$, i.e. $\hat{F}_l(r,{\bi q})=\hat{F}_l({\bi q})\Theta(L-r)$. In this case we can perform an exact matching in $r=L$ comparing the limits ($r \to L^-$, $r \to L^+$) of both $\hat{C}_l$ and $\hat{C}'_l$ (prime means derivative with respect to the variable $r$) deriving from the two expressions (\ref{eq:2}) and (\ref{eq:6}) (see Appendix). The final result is
\begin{equation} \label{eq:8}
\fl\hat{C}_l(r;\ell_q)=\cases{
\phi_{{\bi q},l}\ell_q^{2-\xi}w^{-\nu_0}\left[I_{\nu_l}(w)\displaystyle\int_w^W\rho^{\nu_0+1}K_{\nu_l}(\rho)\rmd\rho\right.&\\\ns
\hspace{2.2cm}\left.+K_{\nu_l}(w)\displaystyle\int_0^w\rho^{\nu_0+1}I_{\nu_l}(\rho)\rmd\rho\right]&for $0<r<L$\\
\phi_{{\bi q},l}\ell_q^{2-\xi}w^{-\nu_0}K_{\nu_l}(w)\displaystyle\int_0^W\rho^{\nu_0+1}I_{\nu_l}(\rho)\rmd\rho&for $L<r<\infty$\;.}
\end{equation}
where $W \equiv w|_{r=L}=2(2-\xi)^{-1}(L/\ell_q)^{(2-\xi)/2}$.\\
Once we assume a small-scale description with respect to the integral scale of the velocity field (i.e. $L_v \to \infty$), the general solution thus depends on three fundamental scales $r$, $L$ and $\ell_q$. These scales represent the separation, the forcing correlation scale and the characteristic length of the inhomogeneities, respectively. From the general solution (\ref{eq:8}) it is important to note (see Appendix) that taking the limit $\ell_q \to \infty$ for fixed $L$ and $r$, (\ref{eq:8}) reduces to the well-known solution for the homogeneous case. Indeed, the latter solution satisfies the equation:
\[r^{-(d-1)}\partial_rr^{d+\xi-1}\partial_r\hat{C}_l-\frac{(d+\xi-1)}{d-1}l(d-2+l)r^{-2}\hat{C}_l+\phi_{{\bi q},l}=0\;.\]
Anyway, in a small scale description with respect to the forcing correlation ($r \ll L$), the presence of a finite $\ell_q$ can reduce the range of pure power-law behaviour because of the presence of the Bessel functions in our solution. This scenario is clearly opposite to the one considered in the homogeneous limit where a pure power-law behaviour is found.\\
In order to get a deeper insight about these two different regimes, in the next section we concentrate ourselves on a purely isotropic situation where the forcing correlation function depends only on $q=|{\bi q}|$ and the correlation of the scalar field coincides with its isotropic sector ($l=0$). This is the simplest, physically relevant assumption with a physical meaning that one can consider to obtain a clear and systematic description of the influence of inhomogeneous contributions.
\section{Analysis for the isotropic case}
\label{sec:3}
We focus here on the isotropic sector ($l=0$) for the two-point equal-time scalar correlation function (the notation $\hat{C}$ is used to indicate $\hat{C}_{l=0}$). The technical advantage now is that alternatively to the method of variation of constants, we can choose the particular solution of (\ref{eq:1}) to be a constant, since the second term in (\ref{eq:1}) vanishes for $l=0$ and the coefficient of the function $\hat{C}(r;\ell_q)$ reduces to a constant. We can repeat the same arguments given for the general case to obtain
\begin{equation} \label{eq:12}
\hat{C}(r;\ell_q)=\cases{
\phi_{\bi q}\ell_q^{2-\xi}+A_2r^{-(d+\xi-2)/2}I_\nu(w)&for $0<r<L$\\
A_3r^{-(d+\xi-2)/2}K_\nu(w)&for $L<r<\infty$\;,}
\end{equation}
where $A_2$,$A_3$ are known functions of $L$ and $\ell_q$ (see Appendix). As discussed in the previous section, the presence of the Bessel functions (the fingerpoint of the scalar inhomogeneities) makes it impossible to see a clear power-law behaviour in the inertial range when $\ell_q/L \sim 1$. This is clearly seen by calculating local-slopes (figure 1) of the difference
\[\hat{S}(r;\ell_q) \equiv \hat{C}(0;\ell_q)-\hat{C}{(r;\ell_q)}\;,\]
strictly related to the second-order structure function, where $\hat{C}(r;\ell_q)$ is the general solution (\ref{eq:12}). For a fixed $L$ and $\xi$ (say, $\xi=4/3$, corresponding to the KOC51 scaling \cite{monyag}) we can change the ratio $\ell_q/L$ and examine the local slope behaviours as a function of $r/L$: the homogeneous case (single power-law behaviour in the inertial range) is recovered only when $\ell_q \gg L$, while for $\ell_q$ of the same order of $L$ a coexistence of power laws spoils the pure scaling of the homogeneous case.
\begin{figure}[h]
\begin{picture}(450,250)
\put(75,0){\includegraphics{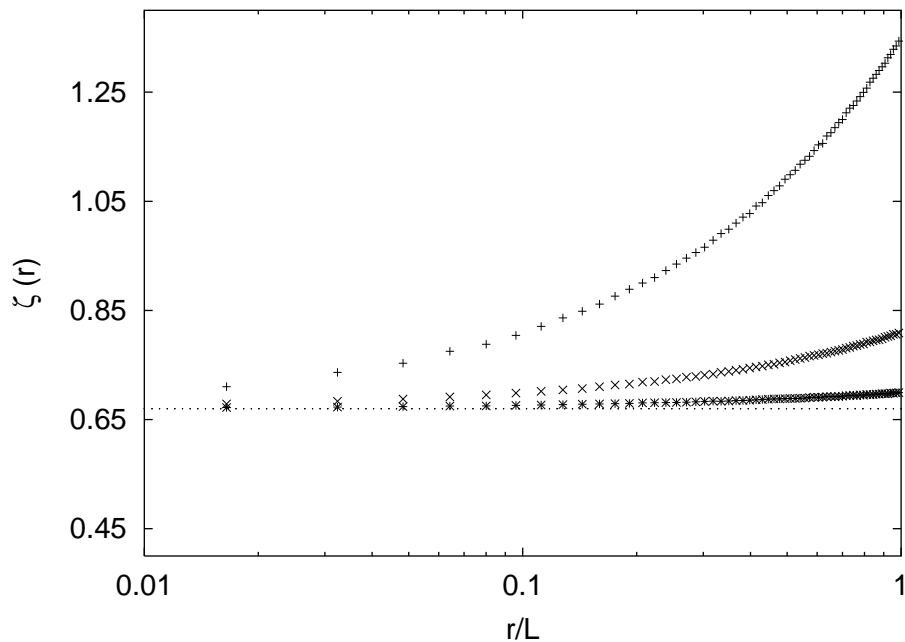}}
\end{picture}
\caption{Local slopes of the correlation function in the inertial range. We plot $\zeta(r)=\frac{\rmd \ln \hat{S} (r;\ell_q)}{\rmd \ln r}$ as a function of $r/L$ for different values of the lenghtscale $\ell_q$. We plot $\ell_q/L=0.1$ ($+$), $\ell_q/L=1$ ($\times$), $\ell_q/L=10$ ($\star$). For comparison, the homogeneous case is also plotted (dotted plot).}
\end{figure}
\\To be more precise, let us focus on the asymptotic properties of (\ref{eq:12}). First of all we can perform the limit $r/\ell_q \to 0$ to obtain
\begin{equation} \label{eq:13}
\hat{C}(r;\ell_q) \approx \hat{C}_{\rm hom}(r;\ell_q) \equiv \cases{
a(\ell_q)+b_2(\ell_q)r^{2-\xi}&for $0<r<L$\\
b_3(\ell_q)r^{2-d-\xi}&for $L<r<\infty$\;,}
\end{equation}
where $a(\ell_q),b_2(\ell_q),b_3(\ell_q)$ can be obtained from the expansion of Bessel functions. In the limit $L \ll \ell_q$, $a$, $b_2$ and $b_3$ reduce to the well-known coefficients of the homogeneous isotropic case $\alpha$,$\beta_2$,$\beta_3$ (see Appendix).\\
In the opposite situation ($\ell_q \ll L$) the function $\hat{C}(r;\ell_q)$ approximates the step function ${\cal C}\Theta(L-r)$, where ${\cal C} \simeq \phi_{\bi q}\ell_q^{2-\xi}$. Thus, if $\phi_{\bi q}=F_0$ is a constant (we shall call it ``forcing of the first kind''), the plot of $\hat{C}(r;\ell_q)$ collapses on the axis of the abscissas when $\ell_q \to 0$. On the contrary, if one wanted to keep ${\cal C}$ finite, a scaling $\phi_{\bi q}=F_0\ell_q^{-(2-\xi)} \propto q^2$ could be assumed: the collapse would now take place for $\ell_q \to \infty$. This simply tells us that some kinds of forcing are not allowed (e.g. we also rule out $\phi_{\bi q}$'s giving unbounded ${\cal C}$'s or $a$'s, as for example $\phi_{\bi q} \sim \ell_q^{\gamma}$ with $\gamma>0$ or $<-(2-\xi)$). The finiteness of both $a$ and ${\cal C}$ may thus be guaranteed assuming e.g. $\phi_{\bi q}=F_{\ell}\ell_q^{-(2-\xi)}+F_LL^{-(2-\xi)}$ (``forcing of the second kind''), with $F_{\ell},F_L$ constants. Of course there would be infinite kinds of allowed forcing, but we will focus on these two because of their physical relevance. In what follows we will consider forcings of the first kind with $F_0=1$ and forcings of the second kind with $F_{\ell}=F_L=1$. The value chosen for $\xi$ is its Kolmogorov value $\xi=4/3$ and we will focus on the three-dimensional case ($d=3$).
\begin{figure}[htbp]
\begin{picture}(450,250)
\put(60,0){\includegraphics{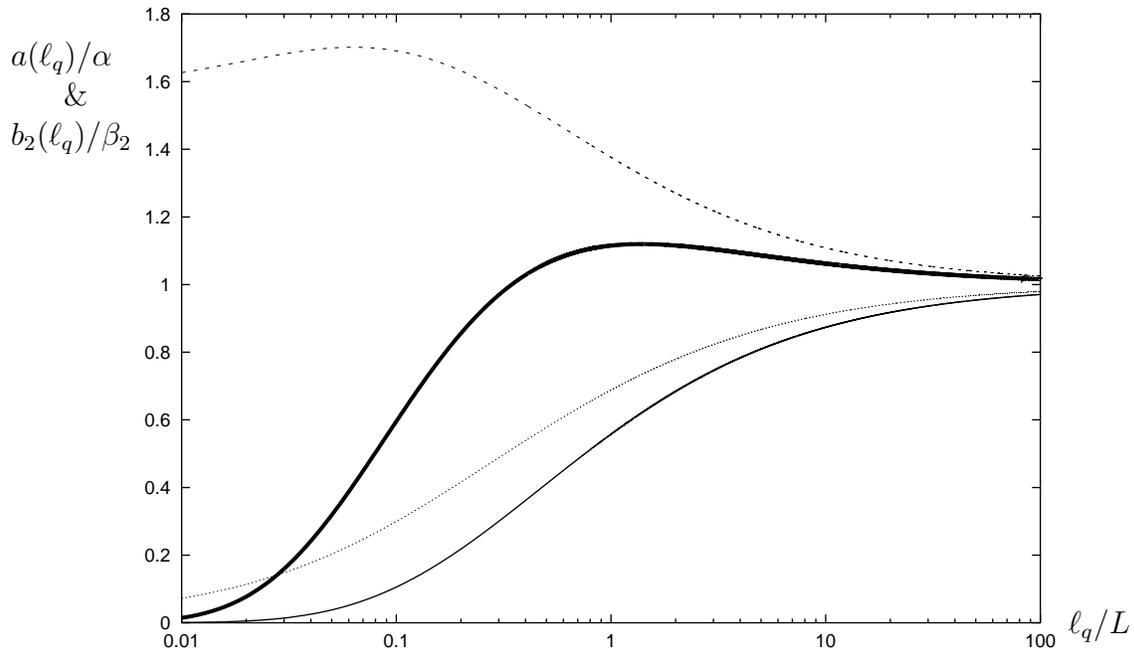}}
\put(20,225){$a(\ell_q)/\alpha$}
\put(40,210){\&}
\put(20,195){$b_2(\ell_q)/\beta_2$}
\put(420,10){$\ell_q/L$}
\end{picture}
\caption{Ratios between ``actual'' (functions of $\ell_q$) and ``homogeneous'' (limit values for $\ell_q \to \infty$) coefficients: dashed lines represent $a(\ell_q)/\alpha$, solid ones $b_2(\ell_q)/\beta$. Thin lines are related to the forcing of the first kind and thick lines are related to the forcing of the second kind.}
\end{figure}
\\Figure 2 represents the ratios of the additive and multiplicative coefficients given by $a(\ell_q)$ and $b_2(\ell_q)$ to the corresponding homogeneous ones $\alpha$ and $\beta_2$, as functions of $\ell_q/L$, for both kinds of forcing. The ``actual'' values attain the ``homogeneous'' ones only for large $\ell_q$'s.
\begin{figure}[htbp]
\begin{picture}(450,250)
\put(60,0){\includegraphics{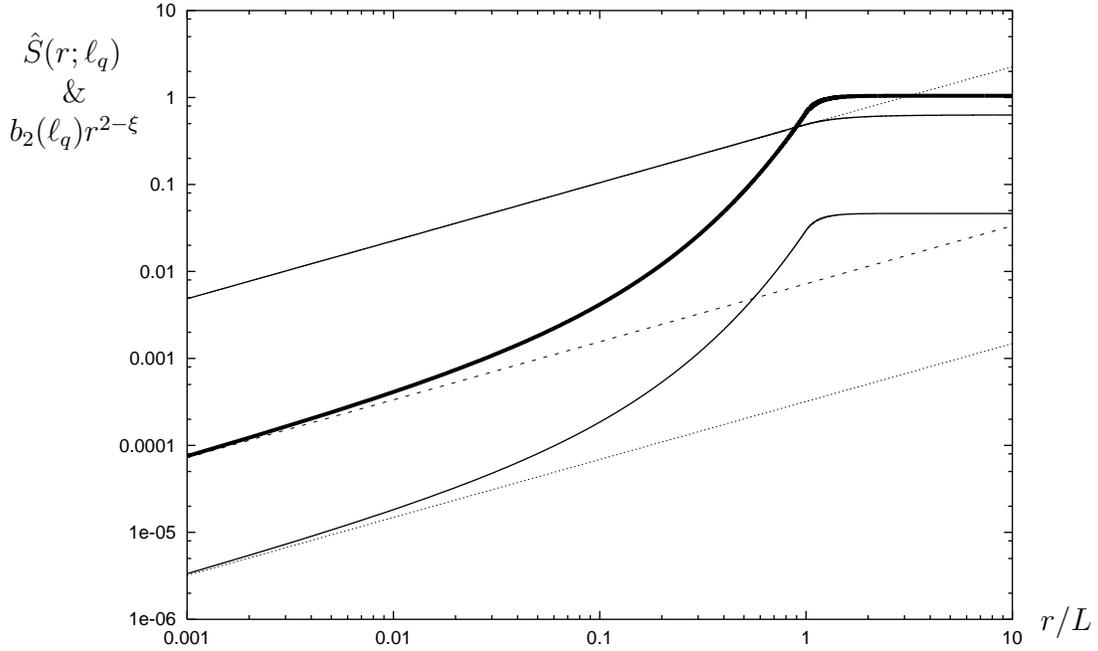}}
\put(35,225){$\hat{S}(r;\ell_q)$}
\put(50,210){\&}
\put(30,195){$b_2(\ell_q)r^{2-\xi}$}
\put(420,10){$r/L$}
\end{picture}
\caption{Plots of $\hat{S}(r;\ell_q)=\hat{C}(0;\ell_q)-\hat{C}(r;\ell_q)$ and of the respective power-law approximations (dashed lines) for $\ell_q/L=10^2$ and forcing of the first kind (upper plot) and $=10^{-2}$ and both kind of forcings (lower plots). Thin lines are related to the forcing of the first kind and thick lines are related to the forcing of the second kind.}
\end{figure}
\\In figure 3 we show the plots of the difference $\hat{C}(0;\ell_q)-\hat{C}(r;\ell_q)$, together with the respective power-law approximations, for two different values of $\ell_q/L$, $10^2$ and $10^{-2}$: in the former case the two kinds of forcing substantially give the same result (only the first kind is thus represented) and the agreement is perfect all over the inertial range, while in the latter the separation takes place for $r<\ell_q$ for both kinds of forcing. One should also notice that by decreasing $\ell_q$, besides the slower convergence to power-law behaviour (as remarked in figure 1), the value $\hat{C}(L;\ell_q)$ tends to decrease with the ``first'' kind of forcing and to increase with the ``second'' kind.
\begin{figure}[htbp]
\begin{picture}(450,250)
\put(60,0){\includegraphics{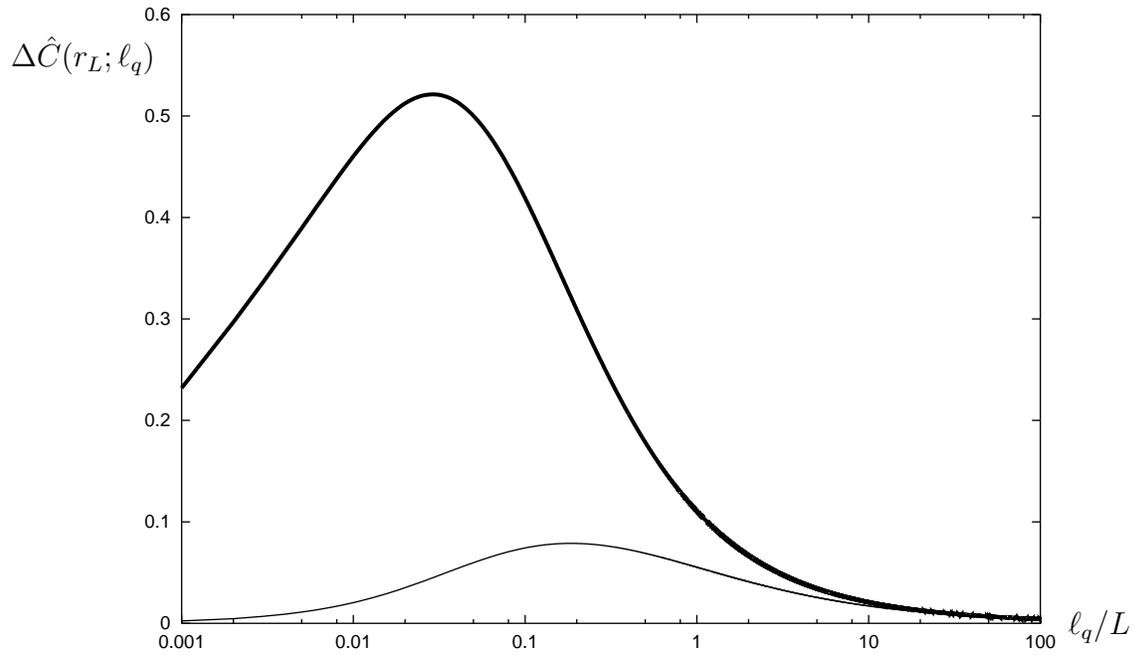}}
\put(20,225){$\Delta\hat{C}(r_L;\ell_q)$}
\put(420,10){$\ell_q/L$}
\end{picture}
\caption{Difference between approximated (power-law behaviour with ``actual'' coefficients $a(\ell_q)$ and $b_2(\ell_q)$) and actual expressions of $\hat{C}(r)$ for $r={\rm golden\ section\ of\ }L$. The thin line is related to the forcing of the first kind while the thick one to the forcing of the second kind.}
\end{figure}
\\Figure 4 shows
\[\Delta\hat{C}(r_L;\ell_q) \equiv \hat{C}_{\rm hom}(r_L;\ell_q)-\hat{C}(r_L;\ell_q)\;,\]
i.e. the difference between the approximated and the actual expressions of $\hat{C}(r;\ell_q)$ as functions of $\ell_q/L$, calculated for $r=r_L$ lying in the inertial range (in this case $r_L$ has been chosen as the golden section of $L$ , but similar plots exist $\forall r<L$). The presence of a maximum is quite intuitive for the first kind of forcing, as both expressions vanish for infinitesimal $\ell_q$, but is remarkable for the second kind, which means that the ``error'' of the approximation becomes negligible not only for large but also for small $\ell_q$'s.\\
The previous discussion has been carried out in the pseudo-Fourier space $({\bi r},{\bi q})$, but the final results must be expressed in the physical space $({\bi r},{\bi z})$ and a superposition is then needed. It is thus useful to analyze some instructive cases of superpositions.
\begin{figure}[htbp]
\begin{picture}(450,250)
\put(60,0){\includegraphics{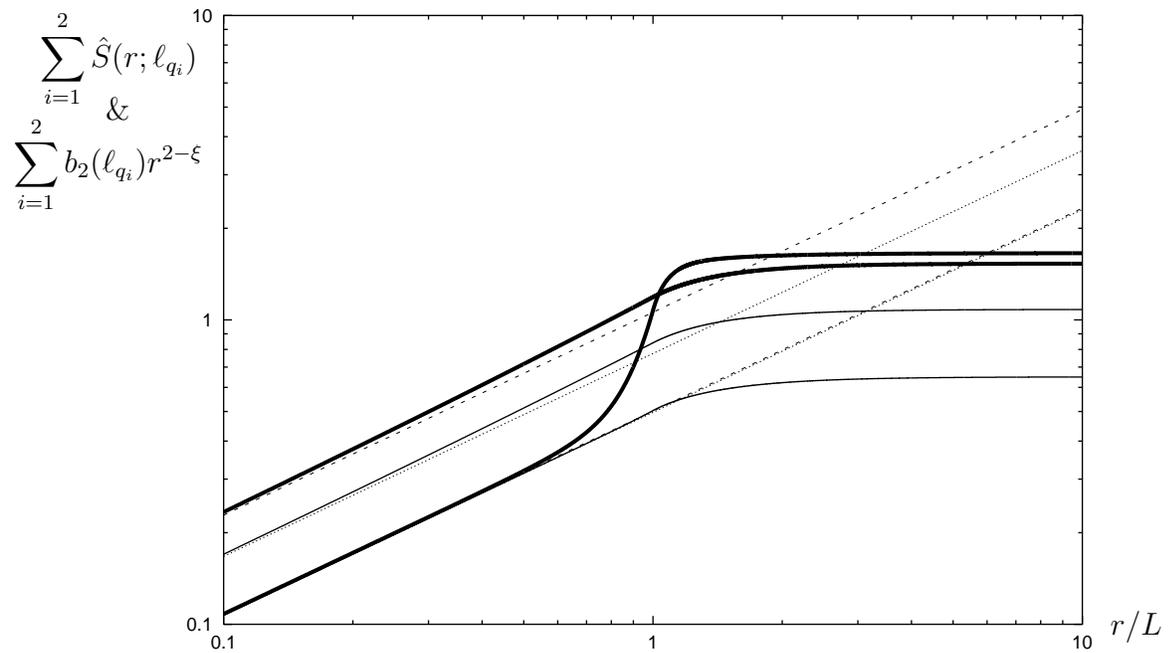}}
\put(15,225){$\displaystyle\sum_{i=1}^2\hat{S}(r;\ell_{q_i})$}
\put(40,205){\&}
\put(5,185){$\displaystyle\sum_{i=1}^2b_2(\ell_{q_i})r^{2-\xi}$}
\put(420,10){$r/L$}
\end{picture}
\caption{Plots of the superpositions of two $\hat{C}(0;\ell_q)-\hat{C}(r;\ell_q)$ and of the respective power-law approximations (dashed lines) for $\ell_{q_1}/L=10^3$ and $\ell_{q_2}/L=1$ (upper two plots) or $=10^{-3}$ (lower two plots). Thin lines are related to the forcing of the first kind and thick lines are related to the forcing of the second kind.}
\end{figure}
\\Figure 5 represents the simplest case of superposition, i.e. the excitation of two $q$'s, e.g. $q_1$ and $q_2$, with constant or $\ell_q$-dependent amplitude. In particular the upper two plots are the (weighted) sum of the modes $\ell_{q_1}/L=10^3$ and $\ell_{q_2}/L=1$ and show very similar behaviours between each other (departure from the $r^{2-\xi}$ straight line at $r$'s about one order of magnitude smaller than the smaller $\ell_q$), while in the lower two $\ell_{q_1}$ is kept fixed but $\ell_{q_2}$ is reduced to $10^{-3}L$. In the former case since the correlation function collapses towards the step function in the limit $\ell_q \to 0 $, the structure function does not ``feel'' the smallest $\ell_q$ in the inertial range. Obviously the restoration of the correct power-law behaviour depends on the degree of convergence of $\hat{C}(r;\ell_q)$ towards the step function (in this case the first kind of forcing converges more rapidly than the second one).
More realistic cases are connected to the excitation of a finite set of discrete modes: in this case the correctness of the power-law approximation is guaranteed (at least) for $r$'s sufficiently smaller than the minimum $\ell_q$. On the contrary, if the forcing has a continuum spectrum, one has to compute the continuum Fourier antitransformed ($\int\rmd^dq\,\rme^{\rmi{\bi q}\cdot{\bi z}} \sim \int\rmd\ell_q\,\ell_q^{-(4-\xi)/2}\ell_q^{-(d-2)(2-\xi)/2}$) which is well defined for the forcings we have considered. 
\section{Conclusions}
\label{sec:4}
The properties of the two-point equal-time correlation function for the Kraichnan model of advection have been studied in presence of anisotropies and inhomogeneities. The system can be described by the following three different scales: the separation ($r$), the forcing correlation length ($L$) and, finally, the lengthscale of the inhomogeneities ($\ell_q$). The model can be treated analytically and the properties of both small scales and large scales can be related to the typical lengthscale $\ell_q$. This offers the possibility to analyze the breaking of translationally invariant properties by means of an external forcing term and to check if the small scale statistics can be regarded as universal in the sense that it does not depend on the details of the inhomogeneous contribution. This somehow universal property is strictly connected to the restoration of a homogeneous limit in an inhomogeneous situation for scales smaller that the typical inhomogeneous one. The homogeneous limit ($r \ll \ell_q$, $L \ll \ell_q $) has been studied and it has been shown how the solution reproduces exactly the one that can be obtained starting from homogeneous equations. On the other side, the homogeneous power-law behaviour is completely spoiled when $\ell_q$ is of the order of the separation $r$ and it can be seen as a ``beating'' of different power laws originating from the scalar inhomogeneities.\\ 
Summarizing, a pure power-law behaviour exists $\forall\ell_q$ going at sufficiently small $r$'s and this is a clear indication of the fact that the statistical description can be seen as the same of the homogeneous case but with a reduced range of pure scaling law behaviour. When we pass to the physical space, and if more inhomogeneous modes are excited, the restoration of an inertial range is guaranteed if the excitation takes place only at large scales or at large scales together with very small scales ($\ell_q \to 0$). The calculations are carried out with the overall exception for those $\ell_q$'s falling in the diffusive range: these have been excluded from our analysis.
\ack
We thank L. Biferale, A. Celani and A. Mazzino for useful discussions and suggestions.
\appendix
\section*{Appendix}
\setcounter{section}{1} 
We start from the equation for the projection of two-point equal-time correlation function $\hat{C}=\hat{C}(r;\ell_q)$ in the anisotropic sector $l$:
\[\fl r^{-(d-1)}\partial_rr^{d+\xi-1}\partial_r\hat{C}_l-\frac{(d+\xi-1)}{d-1}l(d-2+l)r^{-2}\hat{C}_l-\ell_q^{-(2-\xi)}\hat{C}_l+\phi_{{\bi q},l}=0\;.\]
If we assume a forcing whose correlation function is a step function in $r$, i.e. $\hat{F}_{l}(r,{\bi q})=\hat{F}_l({\bi q})\Theta(L-r)$, we can perform an exact matching in $r=L$ by comparing the limits of $\hat{C}_l(r;\ell_q)$ and of $\hat{C}_l'(r;\ell_q)$ (prime means derivative respect to the variable $r$) deriving from the two expressions (\ref{eq:2}) and (\ref{eq:6}):
\[\left\{\begin{array}{ll}
\displaystyle\lim_{r \to L^-}\hat{C}_l(r;\ell_q)=\hat{C}_{l;{\rm part}}(L;\ell_q)+B_2L^{-(d+\xi-2)/2}I_{\nu_l}(W)&\\
\displaystyle\lim_{r \to L^+}\hat{C}_l(r;\ell_q)=B_3L^{-(d+\xi-2)/2}K_{\nu_l}(W)&\;,
\end{array}\right.\]
\[\fl\left\{\begin{array}{lll}
\displaystyle\lim_{r \to L^-}\hat{C}'_l(r;\ell_q)=&\!\!\!\hat{C}'_{l;{\rm part}}(L;\ell_q)&\\
&\displaystyle\!\!\!+B_2\left[-\frac{d+\xi-2}{2}L^{-(d+\xi)/2}I_{\nu_l}(W)+\ell_q^{-(2-\xi)/2}L^{-(d+2\xi-2)/2}I'_{\nu_l}(W)\right]&\\\ms
\displaystyle\lim_{r \to L^+}\hat{C}'_l(r;\ell_q)=&\!\!\!\displaystyle B_3\left[-\frac{d+\xi-2}{2}L^{-(d+\xi)/2}K_{\nu_l}(W)+\ell_q^{-(2-\xi)/2}L^{-(d+2\xi-2)/2}K'_{\nu_l}(W)\right]&\!.
\end{array}\right.\]
The correlation function must be continuous in $r=L$, and the same is for its first derivative. We can thus write the complete solution (for all $\ell_q$) as
\begin{equation} \label{app}
\fl\hat{C}_{l}(r;\ell_q)=\cases{
\hat{C}_{l,{\rm part}}(r;\ell_q)+B_2r^{-(d+\xi-2)/2}I_{\nu_l}(w)&for $0<r<L$\\
B_3r^{-(d+\xi-2)/2}K_{\nu_l}(w)&for $L<r<\infty$\;,}
\end{equation}
where the two constants are
\[B_2=-A\phi_{{\bi q},l}\int_W^{\infty}\rho^{\nu_0+1}K_{\nu_{l}}(\rho)\rmd\rho\]
and
\[B_3=A\phi_{{\bi q},l}\int_0^W\rho^{\nu_0+1}K_{\nu_l}(\rho)\rmd\rho\;.\]
Plugging the values of $B_2$,$B_3$ in (\ref{app}) we can obtain the exact solution written in terms of $w$, $W$ and $\ell_q$:
\[\fl\hat{C}_{l}(r;\ell_q)=\cases{
\phi_{{\bi q},l}\ell_q^{2-\xi}w^{-\nu_0}\left[I_{\nu_l}(w)\displaystyle\int_w^W\rho^{\nu_0+1}K_{\nu_l}(\rho)\rmd\rho\right.&\\\ns
\hspace{2.2cm}\left.+K_{\nu_l}(w)\displaystyle\int_0^w\rho^{\nu_0+1}I_{\nu_l}(\rho)\rmd\rho\right]&for $0<r<L$\\
\phi_{{\bi q},l}\ell_q^{2-\xi}w^{-\nu_0}K_{\nu_l}(w)\displaystyle\int_0^W\rho^{\nu_0+1}I_{\nu_l}(\rho)\rmd\rho&for $L<r<\infty$\;.}\]
In the limit $\ell_q \to \infty$ for fixed $L$ and $r$, we find the well-known solution for the homogeneous case. Indeed, in this limit, for $r<L$ we have
\[\fl\phi_{{\bi q},l}\ell_q^{2-\xi}w^{-\nu_0}I_{\nu_l}(w)\displaystyle\int_w^W\rho^{\nu_0+1}K_{\nu_l}(\rho)\rmd\rho \overset{\ell_q \to \infty}{\longrightarrow} \frac{2\phi_{{\bi q},l}}{(2-\xi)^{2}(\nu_0-\nu_l+2)\nu_l}\left(r^{\zeta^+_l}L^{2-\xi-\zeta^+_l}-r^{2-\xi}\right)\]
\[\fl\phi_{{\bi q},l}\ell_q^{2-\xi}w^{-\nu_0}K_{\nu_l}(w)\displaystyle\int_0^w\rho^{\nu_0+1}I_{\nu_l}(\rho)\rmd\rho \overset{\ell_q \to \infty}{\longrightarrow} \frac{2\phi_{{\bi q},l}}{(2-\xi)^{2}(\nu_0+\nu_l+2)\nu_l}r^{2-\xi}\;,\]
and, for $r>L$,
\[\fl\phi_{{\bi q},l}\ell_q^{2-\xi}w^{-\nu_0}K_{\nu_l}(w)\displaystyle\int_0^W\rho^{\nu_0+1}I_{\nu_l}(\rho)\rmd\rho \overset{\ell_q \to \infty}{\longrightarrow} \frac{2\phi_{{\bi q},l}}{(2-\xi)^{2}(\nu_0+\nu_l+2)\nu_l}r^{\zeta^-_l}L^{2-\xi-\zeta_l^-}\;,\]
where $\zeta_l^-=(-\nu_0-\nu_l)(2-\xi)/2$ and $\zeta_l^+=(-\nu_0+\nu_l)(2-\xi)/2$. We finally obtain
\[\fl\hat{C}_{l}(r)=\cases{ 
\displaystyle\frac{2\phi_{{\bi q},l}}{(2-\xi)^2(\nu_0-\nu_l+2)\nu_l}r^{\zeta^+_l}L^{2-\xi-\zeta^+_l}&\\\ns
\hspace{3cm}\displaystyle-\frac{4\phi_{{\bi q},l}}{(2-\xi)^{2}((\nu_0+2)^2-\nu_l^2)}r^{2-\xi}&for $0<r<L$\\
\displaystyle\frac{2\phi_{{\bi q},l}}{(2-\xi)^{2}(\nu_0+\nu_l+2)\nu_l}r^{\zeta^-_l}L^{2-\xi-\zeta_l^-}&for $L<r<\infty$\;,}\]
which is the solution that can be exactly obained from the homogeneous equation projected along the anisotropic sector $l$:
\[r^{-(d-1)}\partial_rr^{d+\xi-1}\partial_r\hat{C}_l-\frac{(d+\xi-1)}{d-1}l(d-2+l)r^{-2}\hat{C}_l+\phi_{{\bi q},l}=0\;\]
with a step-like forcing term $\phi_{{\bi q},l}$.\\ 
In the isotropic case the particular solution of the forced equation can be chosen as a constant, since the second term in (\ref{eq:1}) vanishes for $l=0$ and the coefficient of the function $\hat{C}(r;\ell_q)$ reduces to a constant. The complete solution can thus be written as (see (\ref{eq:12}))
\[\hat{C}(r;\ell_q)=\cases{
\phi_{\bi q}\ell_q^{2-\xi}+A_2r^{-(d+\xi-2)/2}I_\nu(w)&for $0<r<L$\\
A_3r^{-(d+\xi-2)/2}K_\nu(w)&for $L<r<\infty$\;,}\]
where $\nu$ stands for $\nu_{l=0}$. The two constants are
\[A_2=\frac{\phi_{\bi q}\ell_q^{2-\xi}L^{(d+\xi-2)/2}}{\displaystyle K_\nu(W)\frac{\cal I}{\cal K}-I_\nu(W)}\]
and
\[A_3=\frac{\phi_{\bi q}\ell_q^{2-\xi}L^{(d+\xi-2)/2}}{\displaystyle K_\nu(W)-I_\nu(W)\frac{\cal K}{\cal I}}\;,\]
where
\[{\cal I}=\left[-\frac{d+\xi-2}{2}L^{-(d+\xi)/2}I_\nu(W)+\ell_q^{-(2-\xi)/2}L^{-(d+2\xi-2)/2}I'_\nu(W)\right]\;,\]
\[{\cal K}=\left[-\frac{d+\xi-2}{2}L^{-(d+\xi)/2}K_\nu(W)+\ell_q^{-(2-\xi)/2}L^{-(d+2\xi-2)/2}K'_\nu(W)\right]\;.\]
Performing the limit $r,L \ll \ell_q$ and exploiting the expansion of Bessel functions, we easily obtain
\[\hat{C}(r;\ell_q) \approx \hat{C}_{\rm hom}(r;\ell_q) \equiv \cases{
a(\ell_q)+b_2(\ell_q)r^{2-\xi}&for $0<r<L$\\
b_3(\ell_q)r^{2-d-\xi}&for $L<r<\infty$\;,}\]
where the coefficients can be found after simple but lengthy algebra:
\[a(\ell_q)=\phi_{\bi q}\ell_q^{2-\xi}+A_2\frac{(2-\xi)^{-(d+\xi-2)/(2-\xi)}}{\Gamma(\nu+1)}\ell_q^{-(d+\xi-2)/2}\]
\[b_2(\ell_q)=A_2\frac{(2-\xi)^{-(d-\xi+2)/(2-\xi)}}{\Gamma(\nu+2)}\ell_q^{-(d-\xi+2)/2}\]
\[b_3(\ell_q)=A_3 \frac{(2-\xi)^{(d+\xi-2)/2}\Gamma(\nu)}{2}\ell_q^{(d+\xi-2)/2}\;,\]
with
\[a(\ell_q) \overset{\ell_q \to \infty}{\longrightarrow} \alpha=\frac{\hat{F}(0,{\bf 0})L^{2-\xi}}{(d-1)D_1(2-\xi)(d+\xi-2)}\]
\[b_2(\ell_q) \overset{\ell_q \to \infty}{\longrightarrow} \beta_2=-\frac{\hat{F}(0,{\bf 0})}{d(d-1)D_1(2-\xi)}\]
\[b_3(\ell_q) \overset{\ell_q \to \infty}{\longrightarrow} \beta_3=\frac{\hat{F}(0,{\bf 0})L^d}{d(d-1)D_1(d+\xi-2)}\]
that correspond to the well-known homogeneous isotropic case.

\end{document}